# Feminist epistemology for machine learning systems design

Goda Klumbytė

Department of Participatory IT Design, Faculty of Electrical Engineering and Computer Science, University of Kassel, goda.klumbyte@uni-kassel.de

Hannah Piehl

Department of Participatory IT Design, Faculty of Electrical Engineering and Computer Science, University of Kassel, hannah.piehl@stud.uni-frankfurt.de

Claude Draude

Department of Participatory IT Design, Faculty of Electrical Engineering and Computer Science, University of Kassel, claude.draude@uni-kassel.de

This workshop paper presents a series of feminist epistemological concepts as tools for developing critical, more accountable, and contextualised approaches to machine learning systems design. Namely, we suggest that the methods of situated knowledges or situating, figurations or figuring, diffraction or diffracting, and critical fabulation or speculation can be productively actualised in the field of machine learning systems design. We also suggest that the meta-method for doing this actualisation requires not so much translation but transposition – a creative and critical adaptation to speak to machine learning contexts.

**ACM Reference Format:**
Goda Klumbytė, Hannah Piehl, and Claude Draude. 2023. Feminist epistemology for machine learning systems design. Workshop: A Toolbox of Feminist Wonder: Theories and methods that can make a difference, Conference on Computer-Supported Cooperative Work and Social Computing CSCW '23, October 14-18, 2023, Minneapolis, MN, USA, 5 pages.

## 1 INTRODUCTION

Feminist approaches and concepts have been increasingly used in HCI [1, 2, 3, 4]. With focus on fairness, accountability, transparency and ethics in and of AI/machine learning (ML) systems and their design and emergence of human-centred AI/ML paradigms [5, 6, 7], feminist concepts are gaining ground in AI/ML systems design as well [8, 9, 10]. Here we present our experience of working with feminist epistemology and specifically the methods of situated knowledges, figuration, diffraction, and critical fabulation/speculation to develop more grounded, contextualised, and accountable approaches to ML systems design. We argue that a meta-method for doing such interdisciplinary work is transposition [11,

12]. In music, transposition refers to a variation of melody produced in a different register. In feminist ethics [11] and in our work, it deals with finding creative and critical ways to actualise critical concepts from feminist theory for a context of ML systems design. We first introduce the concepts that we worked with and then describe the main outcomes of such work.

## 2  SITUATED KNOWLEDGES, FIGURATIONS, DIFFRACTION, AND CRITICAL FABULATION

The method of *situating* stems from the concept of situated knowledges and draws on Black feminist scholarship [13] as well as feminist epistemological concerns [14, 15, 16]. It describes how knowledge cannot be seen as universal, objective, or neutral, but rather is partial and always embedded in its contexts [14, 17], and structured by various intersecting positionalities [13]. Closely related to standpoint theory [18, 19, 20], the concept of situated knowledges suggests that (scientific and technological) knowledge is to be understood with regard to the embodied and embedded perspectives it was produced in. Through acknowledging this partiality and grounding different perspectives in their broader technological, social, political, cultural context, one is able to situate knowledge and foster response-ability [13, 14]. Response-ability, defined as the capacity to respond [21, 22], can entail highlighting alternative modes of knowing along with focusing on marginalised viewpoints and experimental modes of knowledge production [13, 23].

*Figuring* as a method is built on the concept of figurations, which are mappings that account for material, political and historical locations [24, 25]. Also referred to as "conceptual personae", they are conceptual figures that embed practices, meanings, notions of power and political controversy [21]. As knowledge production and technological development are intertwined with power relations and cultural meanings, figurations act as tools to conceptualise these intertwinings and problematise technological imaginaries. For example, Donna Haraway's rendering of cyborg as a feminist figuration [26] highlights the entanglement of supposed binaries of humans and machines, as well as technology and society/politics. Figurations encompass imagining and/or identifying figures and envisioning design practices that facilitate analysing such binaries to account for power relations and historical and cultural context of technologies [27].

*Diffraction* is an experimental mode of knowing that combines trans-disciplinary research practices with feminist epistemology [28, 29]. It entails "reading" different bodies of thought "through" another, much like two waves producing diffraction patterns. Borrowed from physics, the term was introduced as a feminist methodology by Karen Barad, who used diffraction to bring together feminist theory and quantum physics to explore notions of agency of matter and socio-material production of meaning [28]. Diffraction also highlights the role of technology and tools of knowledge production in general not only in the way one can research specific phenomena, but also in the way these phenomena get defined [28]. Diffraction has since been used to highlight the material agency of technology and the mutual definition of socio-cultural meanings and technologies [30].

*Critical fabulation* and *critical speculation* is rooted in Black critical and feminist theory [31, 32, 33] and has been employed as a method in critical and speculative design [34, 35, 36, 37]. Fabulation and speculation are used as historically and theoretically informed methods to investigate missing scenarios and reconstruct histories against the dominant narratives. Historian Saidiya Hartman uses this approach to examine archives of transatlantic slavery that exclude the voices of enslaved people and presents critical fabulation as a way of working critically with such silences [38]. As imagination and speculation are key techniques for designing novel technologies, these methods can address (historical) hierarchies and power relations by specifically focusing on missing viewpoints, thus providing a way to counter the "I-methodology" [39] and "white prototypicality" [40, 41, 42] as unquestioned basis of technological design.

## 3 FEMINIST CONCEPTS FOR ML SYSTEMS DESIGN

The concepts of situated knowledges, figurations, diffraction and critical fabulation/speculation were utilised in the context of series of speculative design workshops in 2021, framed as research project and a setting for experimental learning about more inclusive system design at the same time [12, 27]. With the aim of introducing feminist methodologies as tools for design intervention and guidance on how to work with those tools and integrate them into the design process, the workshops tackled one concept per day. After an introduction of the concept, participants engaged with each concept further through exploratory and design exercises, building their speculative ML systems. The detailed process and exercise descriptions are provided in Klumbyte et al. [27] as well as on the website www.critml.org.

Working with the method of situating allowed participants to examine the standpoint of their personal and disciplinary practices and reflect on their accountability. Questioning which communities they felt accountable to and exploring which topics mattered to them acted as a step to further context-sensitive and accountable ML systems [27]. Diffraction helped to contextualise ML systems by investigating the relations between an ML system that participants were designing and its broader structural environments, such as social and institutional contexts, disciplinary and technological background and entrenched cultural norms and values, as well as considering the operational rationale of the ML system [27]. Figurations served as conceptual tools to highlight the cultural, social, and material structure of ML systems and the accountability emerging from locating and uncovering its discursive contexts. Critical fabulation as a method of narrative writing and analysis from different perspectives showed the social embeddedness of stories and opened the design to consider multiple (marginalised) perspectives.

Discussions and productive tensions that emerged in the process of design emphasised that accountability is only attainable if "one is able to call one(self) and systems to account and implicates one(self) in the histories and effects of technologies" [27, pp.8-9]. Furthermore, the notion of response-ability calls for reflection and action. Reflection and positionality were also centred throughout the workshops by a clear focus on power relations and the invitation to align with intersectional feminist and other power-critical approaches. Placing greater focus on the context ML systems design, for example by integrating questions of algorithmic justice and epistemic plurality, encouraged participants to re-orient discussions on fairness towards questions of justice [27]. All of this was a result not only of the workshop design alone but also of the critical valence and orientation introduced by the feminist epistemological concepts themselves.

## 4 DISCUSSION AND CONCLUSION

The four concepts that have been presented above – situated knowledges, figurations, diffraction, and critical fabulation/speculation – are some of the key elements of feminist epistemologies and constitute important feminist methods. What we argue is that working with critical feminist concepts can be challenging in fields such as HCI and systems design. While some literature, particularly when it comes to working with concepts from social sciences and humanities (SSH) in computer science, highlights the need for "translation", "integration", "adaptation" of SSH knowledge for computer science, our proposal would rather be to focus on creating transpositions [11]. In music theory, transposition refers to a process of creating a variation of the main melody. Feminist philosopher Rosi Braidotti meanwhile uses the term to describe the process of "an intertextual, cross-boundary or transversal transfer, in the sense of a leap from one code, field or axis into another, not merely in the quantitative mode of plural multiplications, but rather in the qualitative sense of complex multiplicities" [11, p.6].

For us, transposition is a way of *actualising* feminist theories and methodologies in the field of computer science, and specifically HCI and ML systems design. We argue that this requires a nuanced rendering of specific methods in such a way so as not to lose their critical charge and potential (after all, methods, much like technologies, are also rooted and responsible to their own contexts of creation, their own embedded and embodied perspectives). In the case of the work described here, this was done through constructing a series of exercises that allowed to explore and work with critical concepts as methods for design. We also placed high value on collaborative work and leaving enough space for participants to develop their own interpretations of critical concepts that suit the systems and design settings in question. This collaborative, discussion-rich setting was also important for fostering ethical response-ability of designers. We hope that these concepts can contribute to building a broader feminist toolbox for HCI, systems design, and interaction design that aims to develop accountable, grounded, and reflexive methodologies.

## ACKNOWLEDGMENTS

This research is supported by Volkswagen Foundation grant "Artificial Intelligence and the Society of the Future" as part of the collaborative project "AI Forensics: Accountability through Interpretability in Visual AI Systems".